\documentclass{article}

\usepackage{spconf,amsmath,graphicx}
\usepackage{float}
\usepackage{mathrsfs}
\usepackage{mathtools}
\usepackage{bm}
\usepackage{bbm}
\usepackage{amsthm}
\usepackage{amssymb}
\usepackage{graphicx}

\usepackage{tabularx,booktabs}
\newcolumntype{C}[1]{>{\centering\arraybackslash}p{#1}}
\usepackage{array,multirow}

\makeatletter

\floatstyle{ruled}
\newfloat{algorithm}{tbp}{loa}
\providecommand{\algorithmname}{Algorithm}
\floatname{algorithm}{\protect\algorithmname}

\theoremstyle{plain}

\usepackage{thmtools}

\declaretheoremstyle[
  bodyfont=\normalfont\itshape,
  headformat=\NAME\NUMBER  
]{nospacetheorem}

\@ifundefined{showcaptionsetup}{}{%
 \PassOptionsToPackage{caption=false}{subfig}}
\usepackage{subfig}
\makeatother

\providecommand{\lemmaname}{Lemma}
\providecommand{\propositionname}{Proposition}
\providecommand{\theoremname}{Theorem}

\title{Dictionary Learning with Convex Update (ROMD)}

\name{Cheng Cheng and Wei Dai}

\address{Department of Electrical and Electronic Engineering, Imperial College London, UK}

\begin{document}

\maketitle

\begin{abstract}
Dictionary learning aims to find a dictionary under which the training
data can be sparsely represented, and it is usually achieved by iteratively
applying two stages: sparse coding and dictionary update. 
Typical methods for dictionary update focuses on refining both dictionary atoms and their corresponding sparse coefficients by using the sparsity patterns obtained from sparse coding stage, and hence it is a non-convex bilinear inverse problem.
In this paper, we propose a Rank-One Matrix Decomposition (ROMD) algorithm to recast this challenge into a convex problem by resolving these two variables into a set of rank-one matrices. 
Different from methods in the literature, ROMD updates the whole dictionary at a time using convex programming. The advantages hence include both convergence guarantees for dictionary update and faster convergence of the whole dictionary learning.
The performance of ROMD is compared with other benchmark dictionary learning algorithms. The results show the improvement of ROMD in recovery accuracy, especially in the cases of high sparsity level and fewer observation data.
\end{abstract}

\begin{keywords}
ADMM, conjugate gradient method, convex optimization, dictionary learning
\end{keywords}

\section{Introduction}

Sparse signal representation has drawn extensive research interests in recent decades, and it has found a wide range of applications including signal denoising
\cite{elad2006image,dabov2007image}, restoration \cite{mairal2008sparse,dong2013nonlocally},
source separation \cite{li2006underdetermined,abolghasemi2012blind},
classification \cite{tosic2011dictionary,huang2007sparse}, recognition
\cite{wright2009robust,wright2010sparse,zhang2011sparse}, image super-resolution
\cite{yang2010image,dong2011image} to name a few. The basic idea of sparse signal representation is that an observed signal can be approximated as a linear combination of a few number of codewords selecting from a certain dictionary.  Thus, a resulting topic from this observation namely dictionary learning has drawn numerous researchers' attention. 

As a bilinear inverse problem, typical dictionary learning algorithms alternate between two stages: sparse coding and dictionary update. The principle is to fix one variable and optimize the other. Accordingly, in the sparse coding stage, the purpose is to find the sparse coefficients
based on the fixed dictionary. Its solutions can be generally divided into two categories, greedy algorithms and $\ell_1$-norm relaxation. 
Greedy algorithms include matching pursuit (MP) \cite{mallat1993matching}, orthogonal matching
pursuit (OMP) \cite{pati1993orthogonal,tropp2007signal}, subspace pursuit (SP) \cite{dai2009subspace}, CoSaMP \cite{needell2009cosamp}, that sequentially select the support set from the sparse coefficients. $\ell_1$-norm relaxation, also known as basis pursuit (BP) \cite{chen2001atomic}, convexifies the problem using a surrogate $\ell_1$-norm, the variants of which consist of its unconstrained version named Lasso \cite{tibshirani1996regression} and iterative shrinkage-thresholding algorithms (ISTA) \cite{daubechies2004iterative,hale2007fixed,beck2009fast}.

The other stage dictionary update aims to refine the dictionary using the sparse coefficients obtained from the previous stage. 
In this stage, columns of dictionary, or namely dictionary atoms, are updated either simultaneously \cite{olshausen1996emergence,engan1999method} or sequentially  \cite{aharon2006k,dai2012simultaneous,yu2019bilinear}.
Method of optimal directions (MOD) \cite{engan1999method} is one of the earliest approaches that iteratively alternate between two stages, and the whole dictionary is updated in one step. In the dictionary update stage of MOD, whole sparse coefficient matrix is fixed and then the problem is formulated as a least squares problem.
In many other methods including K-SVD \cite{aharon2006k}, SimCO \cite{dai2012simultaneous} and BLOTLESS \cite{yu2019bilinear}, only the sparsity pattern (the positions of non-zeros) of sparse coefficients is preserved, and both the dictionary and the sparse coefficients are updated. Specifically, K-SVD fixes all but one atom and the corresponding row of sparse coefficients, and obtains their difference to the input signal. Only the elements of the residual at the sparsity pattern is considered, and the dictionary atom and the corresponding sparse coefficients is updated by using singular value decomposition (SVD).
SimCO updates multiple dictionary atoms and the corresponding sparse coefficients by viewing the coefficients as a function of the dictionary and performing a gradient descent with respect to dictionary.
BLOTLESS recasts the dictionary update as a total least squares problem, and updates the blocks of the dictionary and the corresponding elements of sparse coefficients sequentially.
However, these methods update only one atom or a block of the dictionary and the corresponding elements in sparse coefficients at a time, and then sequentially update the whole dictionary.

In the paper, we focus on the dictionary update stage, where only sparse patterns are given. We recast the whole dictionary update problem as a convex optimization programming. 
We decompose the product of two variable into a set of rank-one matrices and use nuclear norm relaxation to promote the low-rankness. By considering the non-smoothness of the nuclear norm, an ADMM framework is applied. 
Numerical tests compare the performance of ROMD with other benchmark dictionary learning algorithms in both synthetic data and real image tests.
The results show the improvement of ROMD in higher recovery accuracy and fewer learning iterations.

\section{Background}

A dictionary learning problem can be formulated as
\begin{align}
\underset{\bm{D},\bm{X}}{\min}\; & \left\Vert \bm{Y}-\bm{D}\bm{X}\right\Vert _{F}^{2}\nonumber \\
{\rm s.t.}\; & \Vert\bm{D}_{:,k}\Vert_{2}=1,\;\Vert\bm{X}_{:,n}\Vert_{0}\le S,\;\forall n\in[N],\forall k\in[K].\label{eq:Chap3-DLOF}
\end{align}
where $\bm{Y}\in\mathbb{R}^{M\times N}$
denotes the observed data, $\bm{D}\in\mathbb{R}^{M\times K}$
represents the unknown dictionary, $\bm{X}\in\mathbb{R}^{K\times N}$
refers to the sparse representation coefficient matrix, $\bm{X}_{:,n}$ is the $n-$th column of $\bm{X}$, $\left\Vert \cdot\right\Vert _{F}$
represents the Frobenius norm, $\left\Vert \cdot\right\Vert _{0}$
indicates the number of non-zeros elements and $[N]:={1,2,\cdots,N}$. 
The constraint of unit norm of the dictionary atom is added to avoid scaling ambiguity, and it is typical that $M<K$, i.e., the dictionary is over-complete. The sparsity
of each column of $\bm{X}$ is assumed at most $S$ such that $S\ll K$. Most
algorithms address (\ref{eq:Chap3-DLOF}) by alternating between two stages: sparse coding and dictionary update.

In sparse coding stage, the dictionary $\bm{D}$ is fixed, and the columns of $\bm{X}$ are obtained by 
\begin{align}
\min_{\bm{X}_{:,n}} & \parallel\bm{Y}_{:,n}-\bm{D}\bm{X}_{:,n}\parallel_{2}^{2}\quad\mathrm{s.t.}\parallel\bm{X}_{:,n}\parallel_{0}\leq S,\:\forall n\in[N],\label{eq:Chap3-sub-problem}
\end{align}
which can be solved by the pursuit algorithms \cite{mallat1993matching,pati1993orthogonal,tropp2007signal,dai2009subspace,needell2009cosamp}.

In dictionary update stage, one can fix either the whole sparse coefficients, e.g. MOD \cite{engan1999method}, or the sparsity patterns, e.g. K-SVD \cite{aharon2006k}, SimCO \cite{dai2012simultaneous}, and Blotless \cite{yu2019bilinear}, to update the dictionary.
In this paper, we focus on the dictionary update problem with given only the sparsity patterns.

\section{Dictionary Learning via ROMD}

\subsection{Problem formulation}

The dictionary update problem can be formulated as
\begin{equation}
\min_{\bm{D},\bm{X}}\parallel\bm{Y}-\bm{D}\bm{X}\parallel_{F}^{2}\quad\mathrm{s.t.}\mathcal{R}_{\Omega}(\bm{X})\neq\bm{0},\label{eq:Chap3-Dictionary-Update-SP}
\end{equation}
where $\Omega$ denotes the support set of $\bm{X}$, and
the operator $\mathcal{R}_{\Omega}(\bm{X})$ retrieves the
values of the entries of $\bm{X}$ in the support set of $\Omega$. Before we introduce ROMD, we first reformulate the matrix multiplication $\bm{D}\bm{X}$ as
\begin{align}
  \bm{D}\bm{X} &= \sum_{k}\bm{D}_{:,k}\bm{X}_{k,:} =\sum_{k}\bm{Q}_{k}^{0}  \nonumber \\
  & = \sum_{k}\mathcal{P}_{k}^{*}(\bm{Q}_{k}), \label{eq:Chap3-reformulating-DX}
\end{align}
where $\bm{D}_{:,k}$, $\bm{X}_{k,:}$ and $\bm{Q}_{k}^{0}$ represent the $k$-th column of $\bm{D}$, the $k$-th row of $\bm{X}$ and the rank-one matrix equals to $\bm{D}_{:,k}\bm{X}_{k,:}$, respectively. Note that a zero entry in $\bm{X}$, say $X_{k,n}$, results in a zero column in $\bm{Q}_k$, i.e., $(\bm{Q}_k)_{:,n} = \bm{D}_{:,k} X_{k,n} = \bm{0}$. To preserve the non-zero columns in $\bm{Q}_{k}^{0}$, we express a projection operator $\mathcal{P}_{k}$ which is formulated as
\begin{align}
\bm{Q}_{k} & =\mathcal{P}_{k}\left(\bm{Q}_{k}^{0}\right)\coloneqq\bm{Q}_{k,:,\Omega_{k}}^{0},   \nonumber\\
\Omega_{k} & \coloneqq\{n:\;\bm{Q}_{k,:,n}^{0}\neq\bm{0}\},\label{eq:defOmega}
\end{align}
and the operator $\mathcal{P}_{k}^{*}$ in \eqref{eq:Chap3-reformulating-DX} denotes the adjoint
operator of $\mathcal{P}_{k}$

According to the rank-one property of the matrix $\bm{Q}_{k}$, we can apply the formulation in \eqref{eq:Chap3-reformulating-DX} to recast
the dictionary update problem \eqref{eq:Chap3-Dictionary-Update-SP} into 
\begin{align}
\min_{\bm{Q}_{k}}\; & \sum_{k}\mathrm{rank}(\bm{Q}_{k})  \nonumber\\
{\rm s.t.}\;  & \bm{Y}= \sum_{k}\mathcal{P}_{k}^{*}\left(\bm{Q}_{k}\right), \label{eq:Chap3-RankMinimization}
\end{align}
which is an optimization problem with respect to a set of matrices $\bm{Q}_{k}$, and is NP-hard regarding to the Rank function. Instead of solving \eqref{eq:Chap3-RankMinimization} directly, we
consider the following convex relaxation problem
\begin{align}
 \min_{\bm{Q}_{k}}\; &  \sum_{k}\left\Vert \bm{Q}_{k}\right\Vert _{*}\label{eq:Chap3-ROMD-formulation}  \nonumber\\
 {\rm s.t.}\; & \bm{Y}= \sum_{k}\mathcal{P}_{k}^{*}\left(\bm{Q}_{k}\right),
\end{align}
by replacing the rank function with nuclear norm $\left\Vert \cdot\right\Vert _{*}$.

It is worth noticing that the bilinear inverse problem \eqref{eq:Chap3-Dictionary-Update-SP} is recast as an optimization problem w.r.t. one variable, which is a set of matrices, in \eqref{eq:Chap3-ROMD-formulation}. Furthermore, ROMD formulation \eqref{eq:Chap3-ROMD-formulation} is convex
and update the whole dictionary simultaneously.

\subsection{An ADMM Solver for ROMD}

As the nuclear norm is non-smooth, we solve the optimization problem (\ref{eq:Chap3-ROMD-formulation}) via
alternating direction method of multipliers (ADMM) \cite{boyd2011distributed}.
For \eqref{eq:Chap3-ROMD-formulation} involves $K$ variables $\bm{Q}_{k}$'s,  we introduce another $K$ auxiliary variables $\bm{Z}_{k}\in\mathbb{R}^{M \times N}$ and rewrite \eqref{eq:Chap3-ROMD-formulation} into the standard ADMM form as
\begin{align}
\min_{\bm{Q}_{k},\bm{Z}_{k}}\;   \sum_{k}\left\Vert \bm{Z}_{k}\right\Vert _{*}\quad \mathrm{s.t.}\;  \bm{A}\bm{x}+\bm{B}\bm{z}=\bm{c},\label{eq:Chap3-standardadmm}
\end{align}
where $\bm{x}=\mathrm{vec}([\bm{Q}_{1},\cdots,\bm{Q}_{K}])\in\mathbb{R}^{M\sum_{k}n_{k}}$ and
$\bm{z}=\mathrm{vec}([\bm{Z}_{1},$ $\cdots,\bm{Z}_{K}])\in\mathbb{R}^{M\sum_{k}n_{k}}$
are the vectors formed by stacking the columns of matrices $\bm{Q}_{k}$
and $\bm{Z}_{k}$ respectively, and $n_{k}$ is the number of elements in $k$-th sparsity patterns $\Omega_{k}$. The matrices $\bm{A}$, $\bm{B}$ and the vector $\bm{c}$ are in the form of
\[
\bm{A}\!\!=\!\!\left[\begin{array}{c}
\bm{I}_{\Omega_{1}}^{T}\otimes\bm{I}_{M}\!\cdots\!\bm{I}_{\Omega_{K}}^{T}\otimes\bm{I}_{M}\\
\bm{I}_{M\sum_{K}n_{k}}
\end{array}\right]\!\in\mathbb{R}^{M(N+\sum_{k}n_{k})\times m\sum_{k}n_{k}},
\]

\[
\bm{B}=\left[\begin{array}{c}
\bm{0}\\
-\bm{I}_{M\sum_{k}n_{k}}
\end{array}\right]\in\mathbb{R}^{M(N+\sum_{k}n_{k})\times m\sum_{k}n_{k}},
\]
and

\[
\bm{c}=\left[\begin{array}{c}
\mathrm{vec}(\bm{Y})\\
\bm{0}\\
\vdots\\
\bm{0}
\end{array}\right]\in\mathbb{R}^{M(N+\sum_{k}n_{k})},
\]
where $\otimes$ denotes the Kronecker product, $\bm{I}_{M}\in\mathbb{R}^{M\times M}$
represents the identity matrix and $\bm{I}_{\Omega_{k}}\in\mathbb{R}^{n_{k}\times N}$
expresses the truncated identity matrix by removing all the rows in $\bm{I}_{N}$
indexed by $j\notin\Omega_{k}$.

For readability, instead of using the standard ADMM form \eqref{eq:Chap3-standardadmm}, we write in the equivalent form as
\begin{align}   \label{eq:Chap3-ADMM-EasyForm}
\min_{\bm{Q}_{k},\bm{Z}_{k}}\; & \sum_{k}\left\Vert \bm{Z}_{k}\right\Vert _{*}  \nonumber\\  
{\rm s.t.}\; & \bm{Y}=\sum_{k}\mathcal{P}_{k}^{*}\left(\bm{Q}_{k}\right)\;{\rm and}\,\bm{Z}_{k}=\bm{Q}_{k},\,\forall k\in [K].  
\end{align}
As there are $MN+M\sum_{k}n_k$ many equality constraints in \eqref{eq:Chap3-ADMM-EasyForm}, we denote the corresponding \emph{scaled} Lagrange multipliers (see \cite[\textsection 3.1.1]{boyd2011distributed} for details) by $\bm{\Lambda}_0 \in \mathbb{R}^{M \times N}$ and $\bm{\Lambda}_{k} \in \mathbb{R}^{M \times n_{k}}$, corresponding to the equality constraints $\bm{Y}=\sum_{k}\mathcal{P}_{k}^{*}\left(\bm{Q}_{k}\right)$ and $\bm{Z}_k = \bm{Q}_k$, respectively. Then the augmented Lagrangian is given by  
\begin{align}
& \mathcal{L}_{\rho}\left(\bm{Q}_{k},\bm{Z}_{k},\bm{\Lambda}_{0},\bm{\Lambda}_{k}\right)  \nonumber  \\
= & \sum_{k}\left( \Vert \bm{Z}_{k}\Vert _{*}  +\frac{\rho}{2}\Vert\bm{Q}_{k}-\bm{Z}_{k}+\bm{\Lambda}_{k}\Vert _{F}^{2} \right)  + \frac{\rho}{2}\Vert \sum_{k}\mathcal{P}_{k}^{*}(\bm{Q}_{k})  \nonumber\\
- & \bm{Y}+\bm{\Lambda}_{0}\Vert _{F}^{2} - \frac{\rho}{2}\Vert\bm{\Lambda}_{0}\Vert_{F}^{2} -\frac{\rho}{2}\sum_{k}\Vert\bm{\Lambda}_{k}\Vert_{F}^{2},
\label{eq:Chap3-Augmented-Lagrangian}
\end{align}
where $\rho>0$ is the penalty parameter.
Then the ADMM iterations are given by
\begin{align}
\bm{Q}_{k}^{l+1} & =\underset{\bm{Q}_{k}}{\arg\min}\,\sum_{k}\left\Vert \bm{Q}_{k}-\bm{Z}_{k}^{l}+\bm{\Lambda}_{k}^{l}\right\Vert _{F}^{2}  +\Vert \sum_{k}\mathcal{P}_{k}^{*}\left(\bm{Q}_{k}\right) \label{eq:Chap3-updateXl} \nonumber\\
& -\bm{Y}+\bm{\Lambda}_{0}^{l}\Vert _{F}^{2},   \\
\bm{Z}_{k}^{l+1} & =\underset{\bm{Z}_{k}}{\arg\min}\,\Vert \bm{Z}_{k}\Vert _{*}  +\frac{\rho}{2}\Vert\bm{Q}_{k}^{l+1}-\bm{Z}_{k}+\bm{\Lambda}_{k}^{l}\Vert _{F}^{2},\label{eq:Chap3-updateZl} \\
\bm{\Lambda}_{k}^{l+1} & =\bm{\Lambda}_{k}^{l}+\bm{Q}_{k}^{l+1}-\bm{Z}_{k}^{l+1},\label{eq:Chap3-updateLambdal}\\
\bm{\Lambda}_{0}^{l+1} & =\bm{\Lambda}_{0}^{l}+\sum_{k}\mathcal{P}_{k}^{*}\left(\bm{Q}_{k}^{l+1}\right)-\bm{Y},\label{eq:Chap3-updateLabmda0}
\end{align}
where $l$ denotes the iteration number.

\begin{algorithm}
\caption{Standard conjugate gradient method \cite{nocedal2006conjugate}\label{alg:Chap3-CGM}}

\textbf{Input}: 

\textbf{$\bm{q}^{0}=\bm{0}$},$\bm{r}^{0}=-\bm{A}^{\mathrm{T}}\bm{b}$
and $\bm{p}^{0}=\bm{r}^{0}$.

\textbf{Output:}

$\bm{q}^{l}$.

\textbf{for} $l=0,1,2,\cdots\;\mathrm{until}\;\bm{r}^{l}=\bm{0}$
\textbf{do}
\begin{itemize}
\item $\alpha^{l}=\frac{(\bm{r}^{l})^{T}\bm{r}^{l}}{(\bm{A}\bm{p}^{l})^{T}(\bm{A}\bm{p}^{l})}.$
\item $\bm{q}^{l+1}=\bm{q}^{l}+\alpha^{l}\bm{p}^{l}$.
\item $\bm{r}^{l+1}=\bm{r}^{l}+\alpha^{l}\bm{A}^{\mathrm{T}}(\bm{A}\bm{p}^{l})$.
\item $\beta^{l+1}=\frac{(\bm{r}^{l+1})^{T}\bm{r}^{l+1}}{(\bm{r}^{l})^{T}\bm{r}^{l}}$.
\item $\bm{p}^{l+1}=-\bm{r}^{l+1}+\beta^{l+1}\bm{p}^{l}$.
\end{itemize}
\textbf{end}
\end{algorithm}

The optimization problem \eqref{eq:Chap3-updateXl} is a quadratic programming in the form of $\parallel\bm{A}\bm{x}-\bm{b}\parallel_{2}^{2}$. The vector $\bm{x}$ and the matrix $\bm{A}$ here are the same as in the standard ADMM formulation of ROMD (\ref{eq:Chap3-standardadmm}), and $\bm{b}=\mathrm{vec}([\bm{B}_{0},\bm{B}_{1},\cdots,\bm{B}_{K}])$,
where $\bm{B}_{0}=\bm{Y}-\bm{\Lambda}_{0}^{l}$ and $\bm{B}_{k}=\bm{Z}_{k}^{l}-\bm{\Lambda}_{k}^{l}$.
For this quadratic problem involves a linear mapping of a huge dimension, it takes a long runtime for using fundamental solvers. According to the specific structure in \eqref{eq:Chap3-updateXl}, we develop a conjugate
gradient (CG) method \cite{nocedal2006conjugate} solver to address this problem. The standard algorithm of which can be found in Algorithm \ref{alg:Chap3-CGM}. 
Instead of directly calculating the matrix-vector products, which cost the most computation in the original CG method, we resolve them into efficient calculations based on the structure of the matrix $\bm{A}$.
Specifically, in order to simplify the matrix-vector products $\bm{A}\bm{p}$ and $\bm{A}^{T}(\bm{A}\bm{p})$, we derive the computations based on the structure of matrix $\bm{A}$ as follows.
Let $\bm{b}'=\bm{A}\bm{p}\in\mathbb{R}^{M(\sum_k n_k +N)}$. Define $\bm{B}'\in\mathbb{R}^{M\times (\sum_k n_k +N) }$ such that $\bm{b}'={\rm vec}(\bm{B}')$ and write $\bm{B}'=[\bm{B}_{1}', \cdots, \bm{B}_{K}',\bm{B}_{0}']$, where $\bm{B}_{k}' \in\mathbb{R}^{M\times n_k}$, $\forall k\in [K]$ and $\bm{B}_{0}' \in\mathbb{R}^{M\times N}$. The vector $\bm{p}$ is in the form of $\bm{p}={\rm vec}(\bm{P})={\rm vec}([\bm{P}_1, \cdots, \bm{P}_K)$, where $\bm{P}_k \in\mathbb{R}^{M \times n_k}$. Then, by the definition of $\bm{A}$, $\bm{B}'$ can be obtained by
$\bm{B}_{k}'=\bm{P}_k$, $\forall k \in [K]$, and $\bm{B}_{0}'=\sum_k \mathcal{P}^{*}_{k}(\bm{P}_k)$ which can be calculated efficiently using the following iterative process:
\begin{center}\fbox{\begin{minipage}{6cm}
    $(\bm{B}')^0=\bm{0}$  \newline
    \textbf{for} $k = 1,\cdots,K$ \newline
    \null\hspace{5mm} $(\bm{B}')^{0}_{:,\Omega_{k}}=(\bm{B}')^{0}_{:,\Omega_{k}}+\bm{B}'_{k}$ \newline
    \textbf{end}
\end{minipage}}
\end{center}
To compute $\bm{w}'=\bm{A}^{T}(\bm{A}\bm{p})=\bm{A}^{T}\bm{b}'$, we define $\bm{W}'$ such that $\bm{w}'={\rm vec(\bm{W}')}$ and write $\bm{W}'=[\bm{W}_{1}',\cdots,\bm{W}_{K}']\in\mathbb{R}^{M \times \sum_k n_k}$. Again, according to the definition of $\bm{A}$, it is straightforward to attain that 
\begin{equation}
    \bm{W}_{k}'=\bm{B}_k+\mathcal{P}_{k}(\bm{B}_0)=\bm{B}_k+(\bm{B}_0)_{:,\Omega_k} \nonumber.
\end{equation}

The optimization problem \eqref{eq:Chap3-updateZl} can be addressed by soft-thresholding algorithm. Define $\hat{\bm{Z}}_{k}=\bm{Q}_{k}^{l+1}+\bm{\Lambda}_{k}^{l}$ and its singular value decomposition (SVD) $\hat{\bm{Z}}_{k}=\bm{U}{\rm diag}(\bm{\sigma})\bm{V}^{T}$. Then $\bm{Z}_{k}^{l+1}$ can be obtained by operating soft-thresholding to $\hat{\bm{Z}}_{k}$
\begin{equation}
\bm{Z}_{k}^{l+1}=\sum_{m=1}^{M}\eta\left(\bm{\sigma}_{m};\frac{1}{\rho}\right)\bm{U}_{:,m}\bm{V}_{:,m}^{T}.\label{eq:Chap3-updateZ}
\end{equation}
where $\eta$ denotes the soft-thresholding operator formulated as
$\eta(\bm{\sigma}_{m};\frac{1}{\rho})={\rm max}\left(0,\bm{\sigma}_{m}-\frac{1}{\rho}\right)$.

According to the above analysis, we propose our dictionary learning
algorithm described in Algorithm \ref{alg:Chap3-Nuclear-norm-minimization}.

\begin{algorithm}
\caption{Dictionary learning with convex dictionary update (ROMD) \label{alg:Chap3-Nuclear-norm-minimization}}

\textbf{Input}: 

Measurement $\bm{Y}$ and initial dictionary $\bm{D}^{0}$.

\textbf{Output:}

Dictionary $\hat{\bm{D}}$ and Sparse coefficient $\hat{\bm{X}}$.

\textbf{Repeat }until convergence (stopping rule)
\begin{itemize}
\item Sparse coding stage: Update the sparsity patterns $\Omega_{l}$ in the sparse coefficients
$\bm{X}$.
\item Dictionary learning stage: 

\textbf{Repeat }until $\Vert\sum_{l=1}^{L}\mathcal{P}_{l}^{*}\left(\bm{Q}_{l}\right)-\bm{Y}\Vert_{F}/\Vert\bm{Y}\Vert_{F}\leq \epsilon$
\begin{itemize}
\item Update $\bm{Q}_{l}$ using CG method.

\item Update $\bm{Z}_{l}$ using soft-thresholding.

\item Update $\bm{\Lambda}_{l}$ via (\ref{eq:Chap3-updateLambdal}).

\item Update $\bm{\Lambda}_{0}$ via (\ref{eq:Chap3-updateLabmda0}).
 
\end{itemize}
\end{itemize}
Update dictionary $\hat{\bm{D}}$ and sparse coefficients $\hat{\bm{X}}$:
$\bm{Q}_{l}=\bm{U}\bm{\Sigma}\bm{V}^{T},\ \hat{\bm{D}}_{:,l}=\bm{u}_{1},\ \hat{\bm{X}}_{l,\Omega_{l}}=\sigma_{1}\bm{v}_{1}$.
\end{algorithm}

\subsection{ROMD with noise}

Consider the noisy case, that is the equality constraint in the ROMD formulation \eqref{eq:Chap3-ROMD-formulation} does not hold exactly. Specifically, we assume $\Vert\bm{Y}-\sum_{k}\mathcal{P}_{k}^{*}\left(\bm{Z}_{k}\right)\Vert_F\leq\epsilon$. For the simplicity of composing, we define the indicator function regarding to the noisy power $\epsilon$ as
\begin{equation}
  \mathbbm{1}_{\Vert \cdot \Vert_F \le \epsilon}(\bm{X})
  :=\begin{cases}
    0, & \text{if}~ \Vert \cdot \Vert_F \le \epsilon, \\
    +\infty & \text{otherwise}.
  \end{cases} 
\end{equation}
By introducing another auxiliary variable $\bm{W}\in\mathbb{R}^{M\times N}$, the noisy form of the optimization problem \eqref{eq:Chap3-ADMM-EasyForm} can be formulated as 
\begin{align}   \label{eq:Chap3-ROMD-ADMMform-noisy}
\min_{\bm{Q}_{k},\bm{Z}_{k}}  \; & \sum_{k}\left\Vert \bm{Z}_{k}\right\Vert _{*} +\mathbbm{1}_{\Vert \cdot \Vert_F \le \epsilon}(\bm{Y}-\bm{W})  \nonumber\\  
{\rm s.t.}\; & \bm{W}=\sum_{k}\mathcal{P}_{k}^{*}\left(\bm{Q}_{k}\right)\;{\rm and}\,\bm{Z}_{k}=\bm{Q}_{k},\,\forall k\in [K].  
\end{align}
Then the augmented Lagrangian can be derived as
\begin{align}
& \mathcal{L}_{\rho}\left(\bm{Q}_{k},\bm{Z}_{k},\bm{W},\bm{\Lambda}_{0},\bm{\Lambda}_{k}\right)  \nonumber  \\
= & \sum_{k}\left( \Vert \bm{Z}_{k}\Vert _{*}  +\frac{\rho}{2}\Vert\bm{Q}_{k}-\bm{Z}_{k}+\bm{\Lambda}_{k}\Vert _{F}^{2} \right)  + \mathbbm{1}_{\Vert \cdot \Vert_F \le \epsilon}(\bm{Y}-\bm{W}) \nonumber\\
+ & \frac{\rho}{2}\Vert \sum_{k}\mathcal{P}_{k}^{*}(\bm{Q}_{k}) 
-  \bm{W}+\bm{\Lambda}_{0}\Vert _{F}^{2} - \frac{\rho}{2}\Vert\bm{\Lambda}_{0}\Vert_{F}^{2} -\frac{\rho}{2}\sum_{k}\Vert\bm{\Lambda}_{k}\Vert_{F}^{2},
\label{eq:Chap3-ROMD-AugLag-Noisy}
\end{align}
and the ADMM iterations are hence given by
\begin{align}
\bm{Q}_{k}^{l+1} & =\underset{\bm{Q}_{k}}{\arg\min}\,\sum_{k}\left\Vert \bm{Q}_{k}-\bm{Z}_{k}^{l}+\bm{\Lambda}_{k}^{l}\right\Vert _{F}^{2} \nonumber\\
&+\Vert \sum_{k}\mathcal{P}_{k}^{*}\left(\bm{Q}_{k}\right)  -\bm{W}+\bm{\Lambda}_{0}^{l}\Vert _{F}^{2},   \\
\bm{Z}_{k}^{l+1} & =\underset{\bm{Z}_{k}}{\arg\min}\,\Vert \bm{Z}_{k}\Vert _{*}  +\frac{\rho}{2}\Vert\bm{Q}_{k}^{l+1}-\bm{Z}_{k}+\bm{\Lambda}_{k}^{l}\Vert _{F}^{2},  \\
\bm{W}^{l+1} &=\underset{\bm{W}}{\arg\min}\,\mathbbm{1}_{\Vert \cdot \Vert_F \le \epsilon}(\bm{Y}-\bm{W})+\Vert \sum_{k}\mathcal{P}_{k}^{*}\left(\bm{Q}_{k}^{l+1}\right) \nonumber\\
&-\bm{W}+\bm{\Lambda}_{0}^{l}\Vert _{F}^{2}, \label{eq:Chap3-ROMD-ADMM-Updaing-W}  \\
\bm{\Lambda}_{k}^{l+1} & =\bm{\Lambda}_{k}^{l}+\left(\bm{Q}_{k}^{l+1}-\bm{Z}_{k}^{l+1}\right),\\
\bm{\Lambda}_{0}^{l+1} & =\bm{\Lambda}_{0}^{l}+\left(\sum_{k}\mathcal{P}_{k}^{*}\left(\bm{Q}_{k}^{l+1}\right)-\bm{W}\right),
\end{align}
where $l$ denotes the iteration number. The key difference between the noisy case and the noise-free case is the updating step of $\bm{W}$ in \eqref{eq:Chap3-ROMD-ADMM-Updaing-W}. Define $\hat{\bm{W}}:=\sum_{k}\mathcal{P}_{k}^{*}\left(\bm{Q}_{k}^{l+1}\right)  +\bm{\Lambda}_{0}^{l}$, and the solution of \eqref{eq:Chap3-ROMD-ADMM-Updaing-W} can be straightforwardly obtained by 
\begin{equation}
    \bm{W}^{l+1}=\epsilon\frac{\hat{\bm{W}}-\bm{Y}}{\Vert\hat{\bm{W}}-\bm{Y}\Vert_{F}}+\bm{Y}.
\end{equation}

\section{Numerical tests}

In this section, we first compare the performance of only dictionary update stage using different dictionary learning algorithms. The performance of whole dictionary learning in both noise-free and noisy cases is also tested.

The observation signal $\bm{Y}$ is formed by $\bm{Y}=\Tilde{\bm{D}}\Tilde{\bm{X}}$, where $\Tilde{\bm{D}}\in\mathbb{R}^{M\times K}$ denotes the ground-truth dictionary and $\Tilde{\bm{X}}\in\mathbb{R}^{K\times N}$ represents the ground-truth sparse coefficients. 
The values of entries in dictionary $\Tilde{\bm{D}}$ are generated by independent standard Gaussian distribution, and each dictionary atom $\Tilde{\bm{D}}_{:,k}$ is with unit $\ell_2$-norm, i.e. $\Vert\Tilde{\bm{D}}_{:,k}\Vert_{2}=1$ $\forall k\in[K]$.
Here we assume the number of measurements is $N$ and the number of nonzero coefficients in the $n$-th column of $\bm{X}^0$ is $S_n$.
The index set of the nonzero coefficients are randomly generated from the uniform distribution on $K \choose S_n$ and the values of the nonzero coefficients are independently generated from the standard Gaussian distribution. In our simulations, we set $S_{n} = S \in \mathbb{N}$, $\forall n\in [N]$.

Performance of a dictionary learning algorithm is evaluated by dictionary recovery error. Consider the permutation ambiguity of the trained dictionary and denote the estimated dictionary as $\hat{\bm{D}}$. As the columns of
both ground-true and estimated dictionaries are normalized, the inner product of a ground truth dictionary atom and the corresponding column of a successfully estimated dictionary tends to be $1$.
Thus, to evaluate the dictionary recovery error, we formulate it as
\begin{equation}
\mathrm{Error}\coloneqq\frac{1}{K}\sum_{k=1}^{K}(1-\hat{\bm{D}}_{:,k}^{T}\Tilde{\bm{D}}_{:,i_{k}}),\label{eq:Chap3-DUE}
\end{equation}
where
\begin{align*}
i_{k} & \coloneqq\arg\max_{i\in\mathcal{I}_{k}}(\hat{\bm{D}}_{:,k}^{T}\Tilde{\bm{D}}_{:,i}),\\
\mathcal{I}_{k} & \coloneqq[K]\backslash\{i_{1},\cdots,i_{K-1}\},
\end{align*}
$\hat{\bm{D}}_{:,k}$ denotes the $k$-th column of estimated
dictionary, and $\Tilde{\bm{D}}_{:,i_k}$ represents the $i_{k}$-th
column of ground truth dictionary.

\begin{figure}
\begin{centering}
\subfloat[Phase transition figure of ROMD.]{\begin{centering}
\includegraphics[scale=0.26]{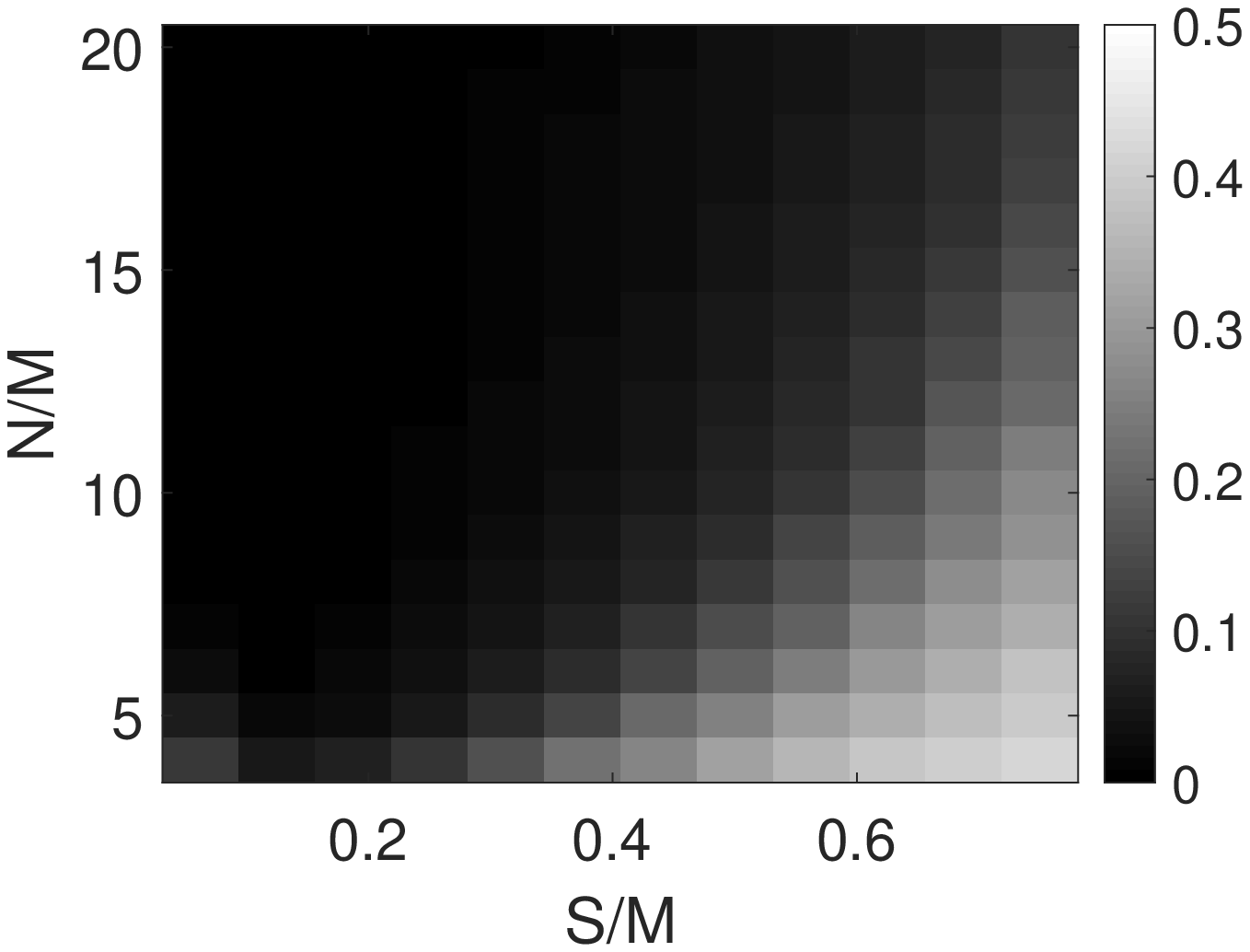}  \label{fig:Chap3-PhsTran-a}
\par\end{centering}
}\\
\subfloat[Phase transition figure of K-SVD.]{\begin{centering}
\includegraphics[scale=0.26]{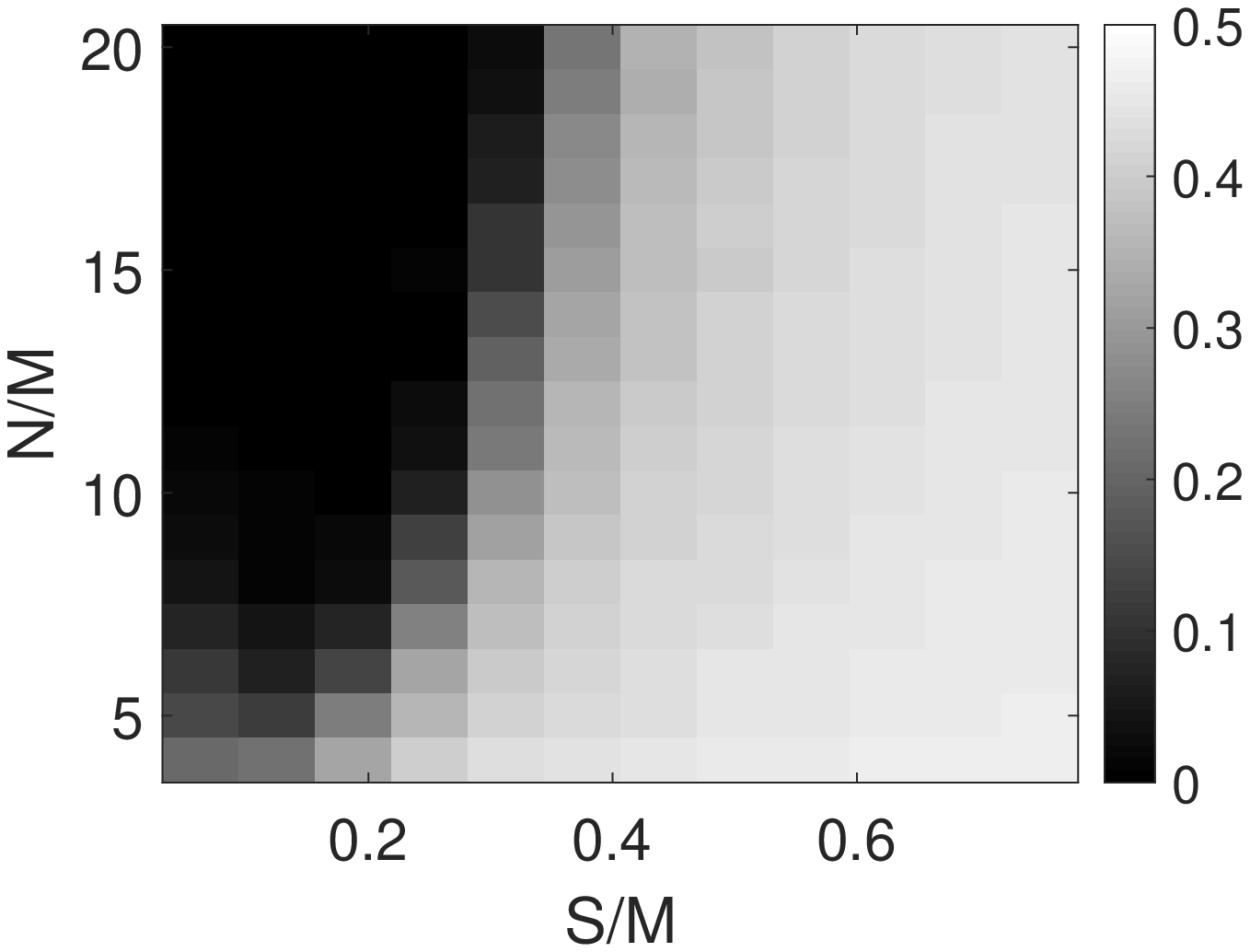}\label{fig:Chap3-PhsTran-b}
\par\end{centering}
}
\subfloat[Phase transition figure of MOD.]{\begin{centering}
\includegraphics[scale=0.26]{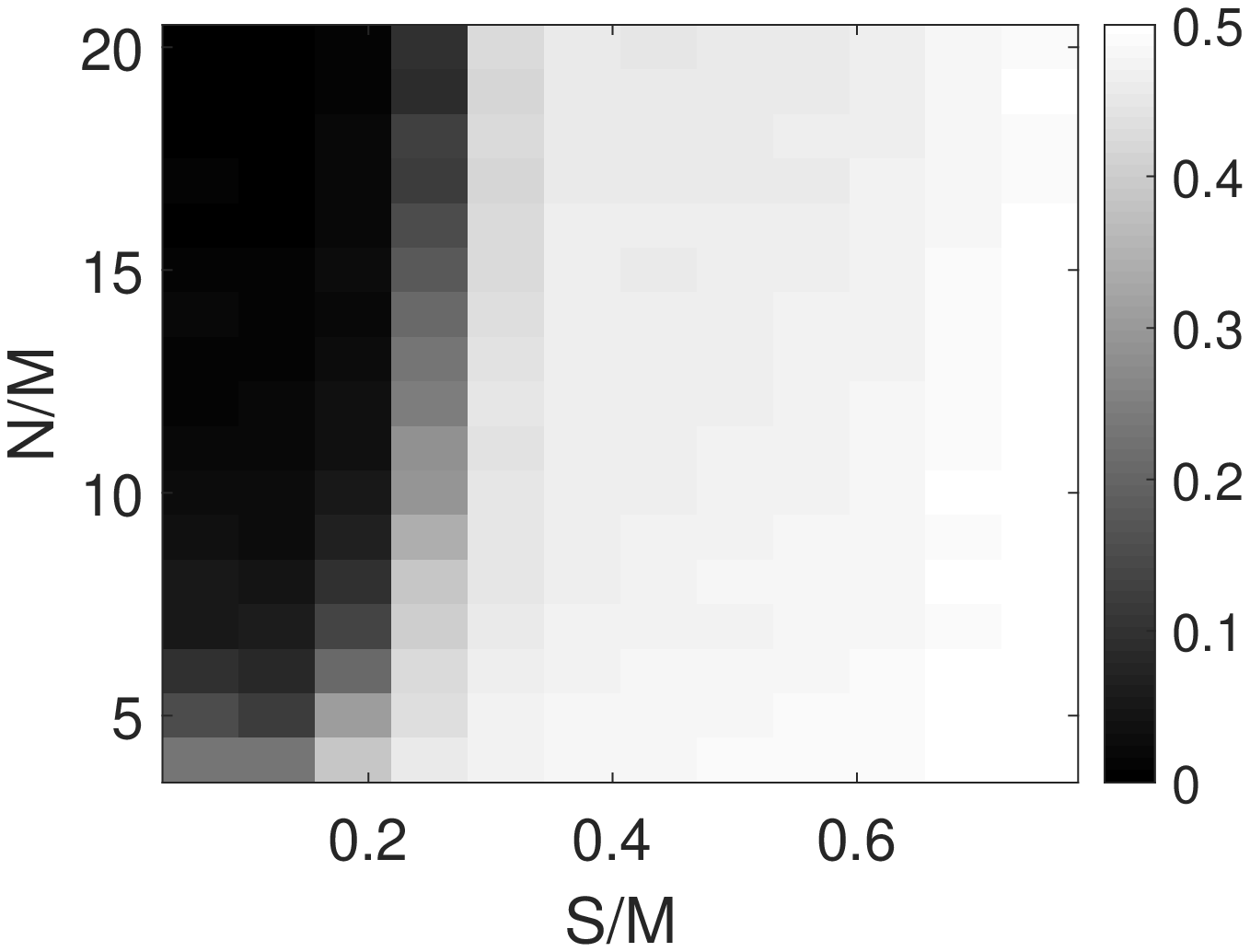} \label{fig:Chap3-PhsTran-c}
\par\end{centering}
}
\par\end{centering}
\caption{\label{fig:Chap3-phase-transition}Comparison of dictionary learning methods with varying sparsity levels and number of measurements. Results are averages of 100 trials.}
\end{figure}

\subsection{Dictionary update test}

In the first test, we only focus on the dictionary update stage, and compare ROMD with two typical dictionary learning algorithms, K-SVD and MOD. We fix the dictionary size to $M=16$ and $K=32$.
To have a more comprehensive view of the performance of these three
algorithms, we vary both
sparsity level ratio $S/M$ and number of samples ratio $N/M$ from $1/16$ to $3/4$ and from $4$ to $20$, respectively.
We repeat each test for 100 trials to acquire an average value.
The phase transition figures of three approaches are depicted in
Figure \ref{fig:Chap3-phase-transition}, and a darker colour represents a lower error. Compared with K-SVD and MOD, ROMD has more dark areas, especially when the sparsity level is high and when the number of samples is low. It is noteworthy that, even for the sparsity level ratio $S/M$ larger than 0.5, ROMD can still reconstruct the whole dictionary with sufficient measurements.
Also noteworthy that, when sparsity level is extremely low, that is $S=1$,
it needs more number of samples compared with the sparsity level $S=2$. The reason is that, for this case , some rows
of sparse coefficient matrix could be all zeros, which eliminates the impact of the corresponding columns of the dictionary.

\subsection{Noise-free dictionary learning test}

\begin{figure}
\begin{centering}
\includegraphics[scale=0.3]{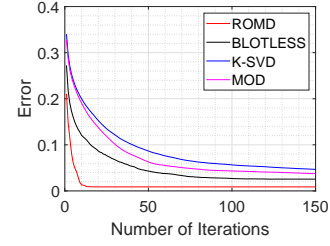}
\par\end{centering}
\centering{}\caption{Comparison of dictionary learning algorithms with settings $M=16,\:K=24,\:N=200$ and $S=3$. Results are averages of 100 trials. }
\label{fig:Chap3-fixed-sample-number}
\end{figure}

Consider the whole dictionary learning process. Here we compare ROMD with BLOTLESS, K-SVD and MOD. OMP is adopted for sparse coding stage for all the algorithms henceforth.
As the whole dictionary learning procedure consists of two stages which alternatively updating two variables, over-tuning the dictionary update stage could cause the whole learning process stuck in a local minimum. To avoid this, we set the stopping criterion of dictionary update stage for ROMD as 
\begin{equation}
\frac{\Vert\sum_k \mathcal{P}^{*}_{k}\bm{Z}_k\Vert_{F}-\Vert\bm{Y}\Vert_{F}}{\Vert\bm{Y}\Vert_{F}}\leq 10^{-5}, \nonumber\end{equation}
which is a value not too small. Here we set the penalty parameter $\rho=0.8$ in ROMD.

Figure \ref{fig:Chap3-fixed-sample-number} illustrates the test where we fix the dictionary size, the sample number and the sparsity level as $M=16$, $K=32$, $N=200$ and $S=3$ respectively, and run each dictionary learning methods for 150 iterations. The results of all the dictionary recovery errors are the averages of 100 trials. Figure  \ref{fig:Chap3-fixed-sample-number} shows the significant improvement of ROMD compared with other benchmark approaches both in the convergence rate and the error rate.
Note that ROMD only needs less than 20 learning iterations to converge, while it takes more than 100 iterations for the other benchmarks.

\begin{figure}
\begin{centering}
\subfloat[$M=$16, $K=$32, $S=$3.]{\begin{centering}
\includegraphics[scale=0.27]{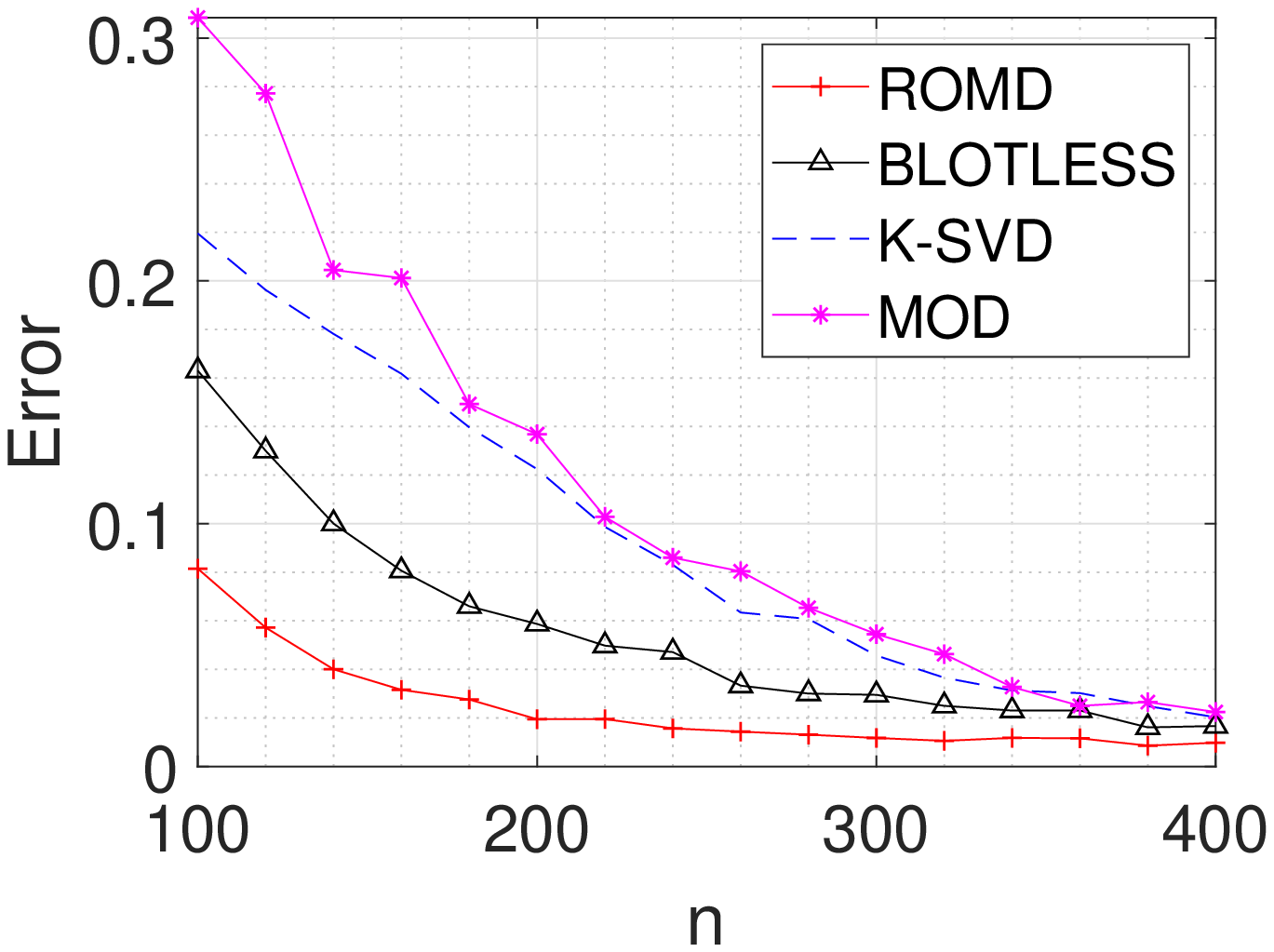}
\par\end{centering}
}
\subfloat[$M=$24, $K=$48, $S=$6]{\begin{centering}
\includegraphics[scale=0.27]{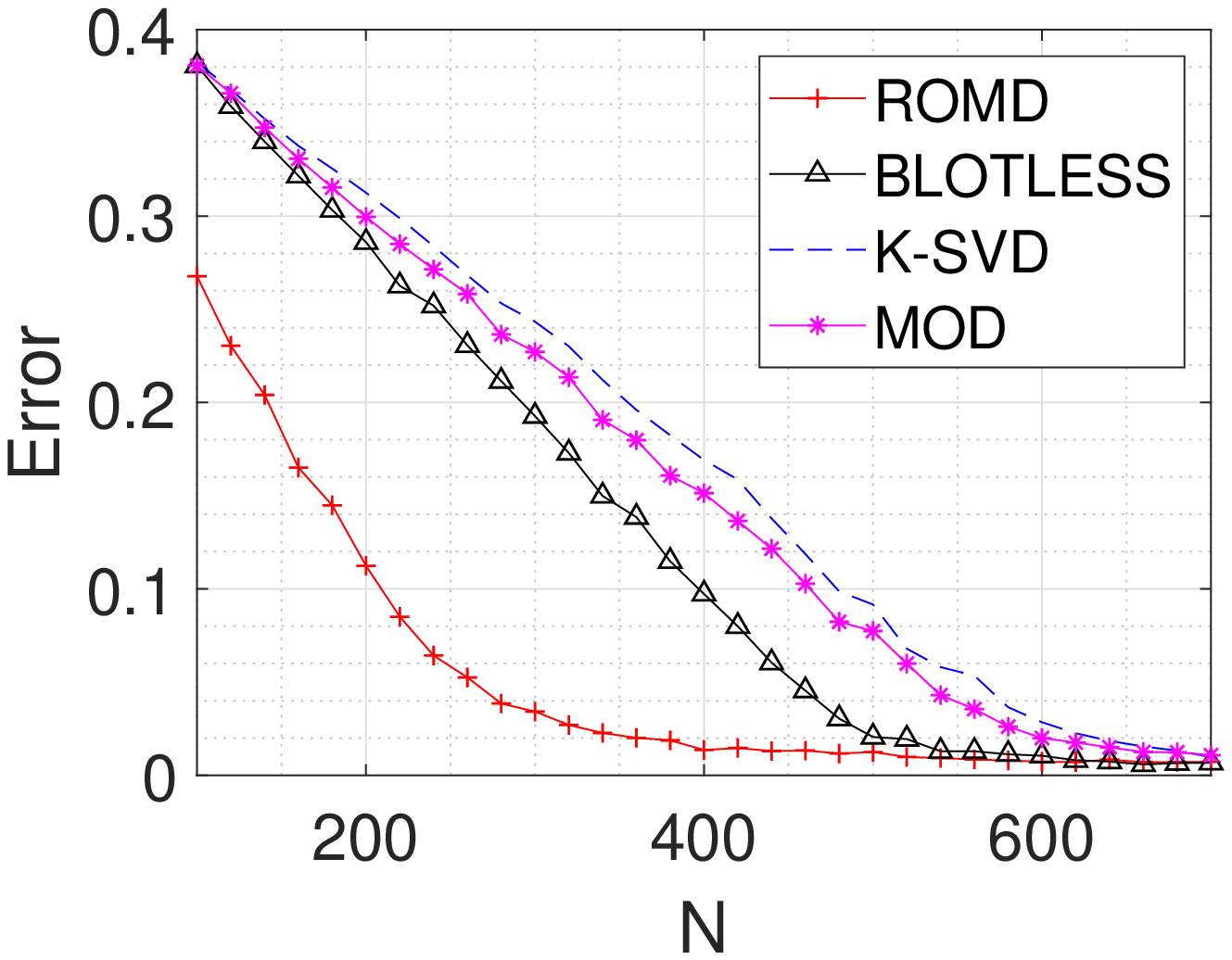}
\par\end{centering}
}
\par\end{centering}
\caption{\label{fig:Chap3-ROMD-syth-test2}Comparison of dictionary learning methods for different number of measurements. Results are averages of 100 trials.}
\end{figure}

Another test compares different dictionary learning methods with varying number of measurements. There are totally two simulations, including two settings of dictionary sizes and sparsity levels of the sparse coefficients.
To  make  every  algorithm  stop at  a  stable  point  even  with  few  samples,  we  set  the  stopping iteration number at 500 for the benchmark algorithms and set 50 iterations for ROMD. Again, we set the penalty parameter $\rho=0.8$ in ROMD for both simulations, and we run 100 trails to get an average value. 
Figure \ref{fig:Chap3-ROMD-syth-test2} demonstrates that for both tests, ROMD requires fewer samples to reconstruct the dictionary and can obtain a lower error level than the other benchmark algorithms.

\subsection{Noisy dictionary learning test}

\begin{figure}
\begin{centering}
\subfloat[$M=$16, $K=$32, $S=$3, SNR$=30$dB.]{\begin{centering}
\includegraphics[scale=0.269]{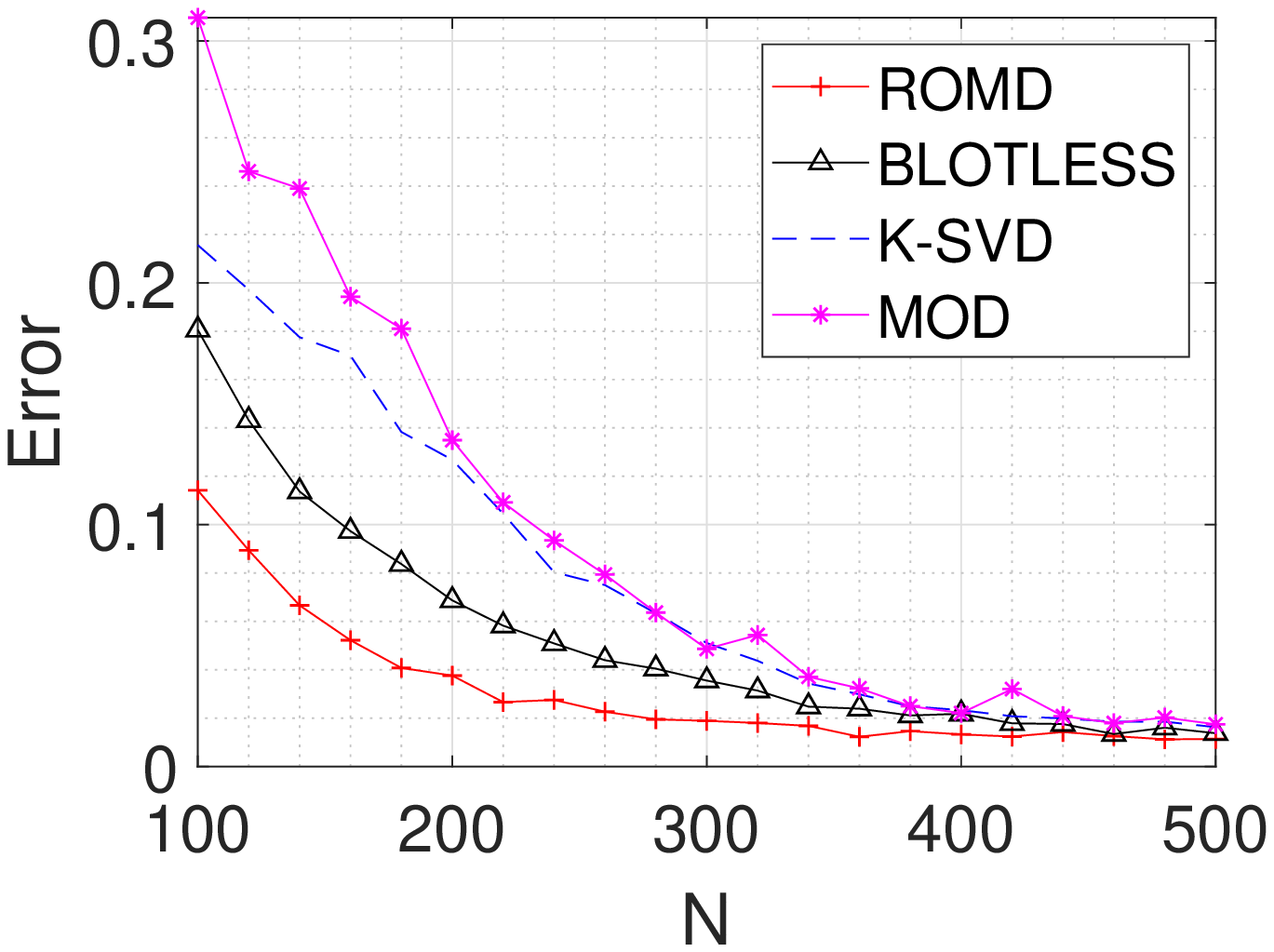}
\par\end{centering}
}\hspace{0.1mm}
\subfloat[$M=16$, $K=32$, $S=3$, SNR$=20$dB.]{\begin{centering}
\includegraphics[scale=0.269]{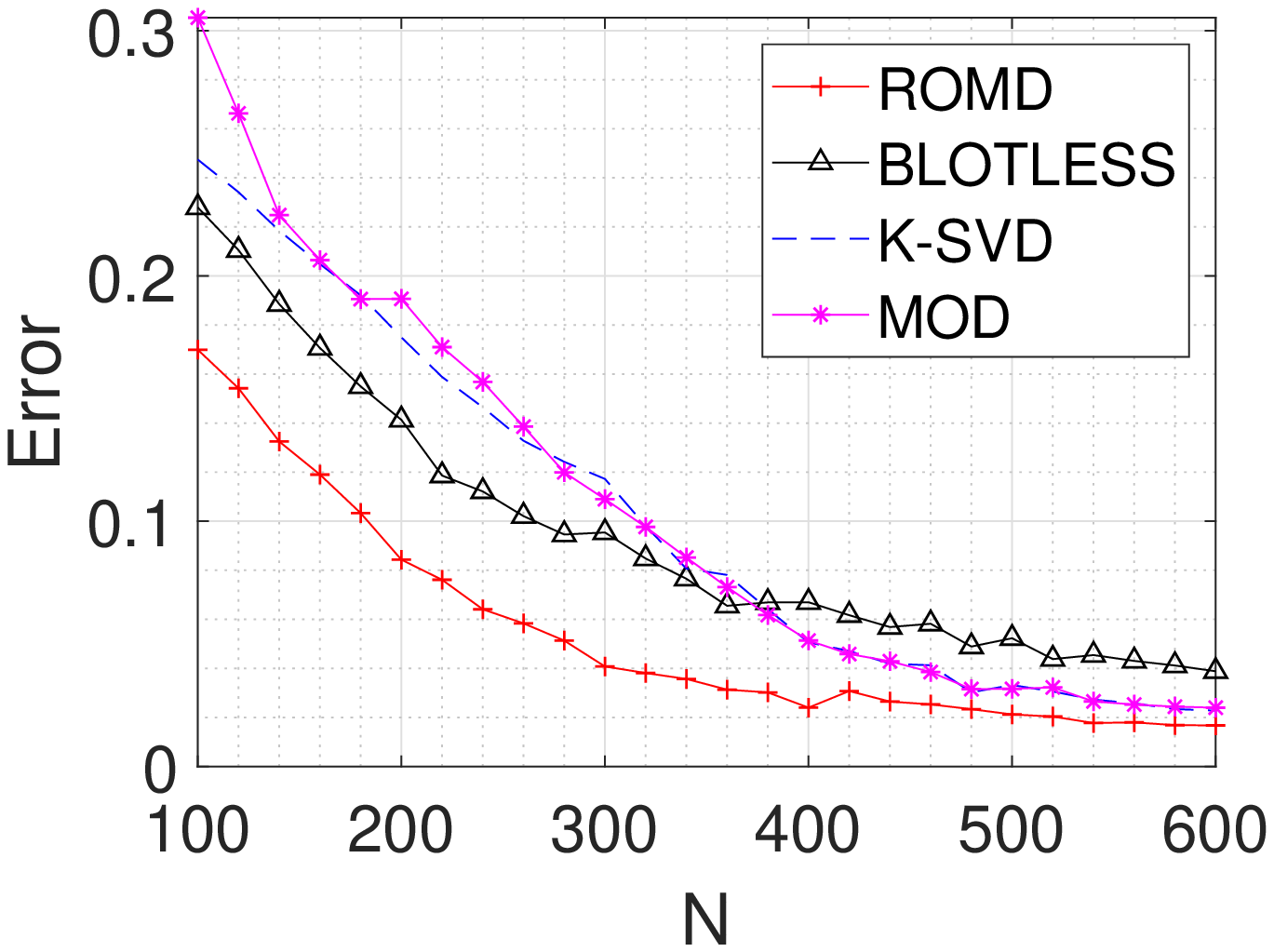}
\par\end{centering}
}
\par\end{centering}
\caption{\label{fig:Chap3-ROMD-Noisy}Comparison of dictionary learning methods for different dictionary sizes, different sparsity levels and different number of measurements. Results are averages of 100 trials.}
\end{figure}

In this test, we compare different dictionary learning methods in noisy cases. We fix the dictionary size and sparsity level as $M=16$, $K=32$, $S=3$ respectively, and we test the noisy cases with SNR$=30$dB and SNR$=20$dB. Here, as there is a noise in observations, we set a relatively small value for penalty parameter $\rho=0.02$ to ensure that the denoised term can be decomposed into rank-one matrices. Again, we vary the number of samples and run each test for 100 trails. The results are illustrated in Figure \ref{fig:Chap3-ROMD-Noisy}, which indicate that ROMD still outperforms the benchmark algorithms.

\section{Conclusion}

In this paper, we propose a dictionary learning algorithm rank-one matrix decomposition (ROMD), that recasts the dictionary update stage into a convex optimization problem by matrix decomposition technique. By using ROMD formulation, the whole dictionary can be updated at a time.
For the non-smoothness of nuclear norm, we apply an ADMM solver. In numerical tests, we compare ROMD with other benchmark algorithms in both noise-free and noisy cases. The results demonstrate that ROMD outperforms the other benchmarks in terms of the fewer learning iterations, fewer required number of samples and lower dictionary recovery error. 

\clearpage

\bibliographystyle{IEEEtran}
\bibliography{RefDU}

\begin{thebibliography}{10}
\providecommand{\url}[1]{#1}
\csname url@samestyle\endcsname
\providecommand{\newblock}{\relax}
\providecommand{\bibinfo}[2]{#2}
\providecommand{\BIBentrySTDinterwordspacing}{\spaceskip=0pt\relax}
\providecommand{\BIBentryALTinterwordstretchfactor}{4}
\providecommand{\BIBentryALTinterwordspacing}{\spaceskip=\fontdimen2\font plus
\BIBentryALTinterwordstretchfactor\fontdimen3\font minus
  \fontdimen4\font\relax}
\providecommand{\BIBforeignlanguage}[2]{{%
\expandafter\ifx\csname l@#1\endcsname\relax
\typeout{** WARNING: IEEEtran.bst: No hyphenation pattern has been}%
\typeout{** loaded for the language `#1'. Using the pattern for}%
\typeout{** the default language instead.}%
\else
\language=\csname l@#1\endcsname
\fi
#2}}
\providecommand{\BIBdecl}{\relax}
\BIBdecl

\bibitem{elad2006image}
M.~Elad and M.~Aharon, ``Image denoising via sparse and redundant
  representations over learned dictionaries,'' \emph{IEEE Transactions on Image
  processing}, vol.~15, no.~12, pp. 3736--3745, 2006.

\bibitem{dabov2007image}
K.~Dabov, A.~Foi, V.~Katkovnik, and K.~Egiazarian, ``Image denoising by sparse
  3-d transform-domain collaborative filtering,'' \emph{IEEE Transactions on
  image processing}, vol.~16, no.~8, pp. 2080--2095, 2007.

\bibitem{mairal2008sparse}
J.~Mairal, M.~Elad, and G.~Sapiro, ``Sparse representation for color image
  restoration,'' \emph{IEEE Transactions on image processing}, vol.~17, no.~1,
  pp. 53--69, 2008.

\bibitem{dong2013nonlocally}
W.~Dong, L.~Zhang, G.~Shi, and X.~Li, ``Nonlocally centralized sparse
  representation for image restoration,'' \emph{IEEE Transactions on Image
  Processing}, vol.~22, no.~4, pp. 1620--1630, 2013.

\bibitem{li2006underdetermined}
Y.~Li, S.-I. Amari, A.~Cichocki, D.~W. Ho, and S.~Xie, ``Underdetermined blind
  source separation based on sparse representation,'' \emph{IEEE Transactions
  on signal processing}, vol.~54, no.~2, pp. 423--437, 2006.

\bibitem{abolghasemi2012blind}
V.~Abolghasemi, S.~Ferdowsi, and S.~Sanei, ``Blind separation of image sources
  via adaptive dictionary learning,'' \emph{IEEE Transactions on Image
  Processing}, vol.~21, no.~6, pp. 2921--2930, 2012.

\bibitem{tosic2011dictionary}
I.~Tosic and P.~Frossard, ``Dictionary learning,'' \emph{IEEE Signal Processing
  Magazine}, vol.~28, no.~2, pp. 27--38, 2011.

\bibitem{huang2007sparse}
K.~Huang and S.~Aviyente, ``Sparse representation for signal classification,''
  in \emph{Advances in neural information processing systems}, 2007, pp.
  609--616.

\bibitem{wright2009robust}
J.~Wright, A.~Y. Yang, A.~Ganesh, S.~S. Sastry, and Y.~Ma, ``Robust face
  recognition via sparse representation,'' \emph{IEEE transactions on pattern
  analysis and machine intelligence}, vol.~31, no.~2, pp. 210--227, 2009.

\bibitem{wright2010sparse}
J.~Wright, Y.~Ma, J.~Mairal, G.~Sapiro, T.~S. Huang, and S.~Yan, ``Sparse
  representation for computer vision and pattern recognition,''
  \emph{Proceedings of the IEEE}, vol.~98, no.~6, pp. 1031--1044, 2010.

\bibitem{zhang2011sparse}
L.~Zhang, M.~Yang, and X.~Feng, ``Sparse representation or collaborative
  representation: Which helps face recognition,'' in \emph{Computer vision
  (ICCV), 2011 IEEE international conference on}.\hskip 1em plus 0.5em minus
  0.4em\relax IEEE, 2011, pp. 471--478.

\bibitem{yang2010image}
J.~Yang, J.~Wright, T.~S. Huang, and Y.~Ma, ``Image super-resolution via sparse
  representation,'' \emph{IEEE transactions on image processing}, vol.~19,
  no.~11, pp. 2861--2873, 2010.

\bibitem{dong2011image}
W.~Dong, L.~Zhang, G.~Shi, and X.~Wu, ``Image deblurring and super-resolution
  by adaptive sparse domain selection and adaptive regularization,'' \emph{IEEE
  Transactions on Image Processing}, vol.~20, no.~7, pp. 1838--1857, 2011.

\bibitem{mallat1993matching}
S.~Mallat and Z.~Zhang, ``Matching pursuit with time-frequency dictionaries,''
  Courant Institute of Mathematical Sciences New York United States, Tech.
  Rep., 1993.

\bibitem{pati1993orthogonal}
\emph{Orthogonal matching pursuit: Recursive function approximation with
  applications to wavelet decomposition}.\hskip 1em plus 0.5em minus
  0.4em\relax IEEE, 1993.

\bibitem{tropp2007signal}
J.~A. Tropp and A.~C. Gilbert, ``Signal recovery from random measurements via
  orthogonal matching pursuit,'' \emph{IEEE Transactions on information
  theory}, vol.~53, no.~12, pp. 4655--4666, 2007.

\bibitem{dai2009subspace}
W.~Dai and O.~Milenkovic, ``Subspace pursuit for compressive sensing signal
  reconstruction,'' \emph{IEEE transactions on Information Theory}, vol.~55,
  no.~5, pp. 2230--2249, 2009.

\bibitem{needell2009cosamp}
D.~Needell and J.~A. Tropp, ``Cosamp: Iterative signal recovery from incomplete
  and inaccurate samples,'' \emph{Applied and computational harmonic analysis},
  vol.~26, no.~3, pp. 301--321, 2009.

\bibitem{chen2001atomic}
S.~S. Chen, D.~L. Donoho, and M.~A. Saunders, ``Atomic decomposition by basis
  pursuit,'' \emph{SIAM review}, vol.~43, no.~1, pp. 129--159, 2001.

\bibitem{tibshirani1996regression}
R.~Tibshirani, ``Regression shrinkage and selection via the lasso,''
  \emph{Journal of the Royal Statistical Society. Series B (Methodological)},
  pp. 267--288, 1996.

\bibitem{daubechies2004iterative}
I.~Daubechies, M.~Defrise, and C.~De~Mol, ``An iterative thresholding algorithm
  for linear inverse problems with a sparsity constraint,''
  \emph{Communications on Pure and Applied Mathematics: A Journal Issued by the
  Courant Institute of Mathematical Sciences}, vol.~57, no.~11, pp. 1413--1457,
  2004.

\bibitem{hale2007fixed}
E.~T. Hale, W.~Yin, and Y.~Zhang, ``A fixed-point continuation method for
  l1-regularized minimization with applications to compressed sensing,''
  \emph{CAAM TR07-07, Rice University}, vol.~43, p.~44, 2007.

\bibitem{beck2009fast}
A.~Beck and M.~Teboulle, ``A fast iterative shrinkage-thresholding algorithm
  for linear inverse problems,'' \emph{SIAM journal on imaging sciences},
  vol.~2, no.~1, pp. 183--202, 2009.

\bibitem{olshausen1996emergence}
B.~A. Olshausen and D.~J. Field, ``Emergence of simple-cell receptive field
  properties by learning a sparse code for natural images,'' \emph{Nature},
  vol. 381, no. 6583, p. 607, 1996.

\bibitem{engan1999method}
K.~Engan, S.~O. Aase, and J.~H. Husoy, ``Method of optimal directions for frame
  design,'' in \emph{Acoustics, Speech, and Signal Processing, 1999.
  Proceedings., 1999 IEEE International Conference on}, vol.~5.\hskip 1em plus
  0.5em minus 0.4em\relax IEEE, 1999, pp. 2443--2446.

\bibitem{aharon2006k}
M.~Aharon, M.~Elad, A.~Bruckstein \emph{et~al.}, ``K-svd: An algorithm for
  designing overcomplete dictionaries for sparse representation,'' \emph{IEEE
  Transactions on signal processing}, vol.~54, no.~11, p. 4311, 2006.

\bibitem{dai2012simultaneous}
W.~Dai, T.~Xu, and W.~Wang, ``Simultaneous codeword optimization (simco) for
  dictionary update and learning,'' \emph{IEEE Transactions on Signal
  Processing}, vol.~60, no.~12, pp. 6340--6353, 2012.

\bibitem{yu2019bilinear}
Q.~Yu, W.~Dai, Z.~Cvetkovic, and J.~Zhu, ``Bilinear dictionary update via
  linear least squares,'' in \emph{ICASSP 2019-2019 IEEE International
  Conference on Acoustics, Speech and Signal Processing (ICASSP)}.\hskip 1em
  plus 0.5em minus 0.4em\relax IEEE, 2019, pp. 7923--7927.

\bibitem{boyd2011distributed}
S.~Boyd, N.~Parikh, E.~Chu, B.~Peleato, J.~Eckstein \emph{et~al.},
  ``Distributed optimization and statistical learning via the alternating
  direction method of multipliers,'' \emph{Foundations and
  Trends{\textregistered} in Machine learning}, vol.~3, no.~1, pp. 1--122,
  2011.

\bibitem{nocedal2006conjugate}
J.~Nocedal and S.~J. Wright, ``Conjugate gradient methods,'' \emph{Numerical
  optimization}, pp. 101--134, 2006.

\end{thebibliography}

\end{document}